\documentclass[aps,prl,floatfix,twocolumn,reprint,amsmath,amssymb,superscriptaddress]{revtex4-1}
\usepackage{amsfonts}
\usepackage{mathrsfs}
\usepackage{amsmath}
\usepackage{color}
\usepackage{natbib}
\usepackage{graphicx}
\usepackage{bm}
\usepackage{amssymb}
\usepackage{xspace}
\usepackage{epstopdf}
\usepackage{dcolumn}
\usepackage{longtable}
\usepackage{multirow}
\usepackage[colorlinks=true, letterpaper=true, pdfstartview=FitV, linkcolor=blue, citecolor=blue, urlcolor=blue]{hyperref}
\UseRawInputEncoding

\makeatletter

\newcommand{\Rmnum}[1]{\expandafter\@slowromancap\romannumeral #1@}
\makeatother

\begin{document}

\title{Hybrid Nodal-chain Semimetal State in MgCaN$_2$}

\author{Hongbo Wu}
\affiliation{College of Physics, Hebei Key Laboratory of Photophysics Research and Application, Hebei Normal University, Shijiazhuang 050024, China}

\author{Da-Shuai Ma}
\affiliation{Institute for Structure and Function $\&$ Department of Physics, Chongqing University, Chongqing 400044, China}

\author{Botao Fu}
\email{fubotao2008@gmail.com}
\affiliation{College of Physics and Electronic Engineering, Center for Computational Sciences, Sichuan Normal University, Chengdu, 610068, China}

\begin{abstract}

The distinct over-tilting of band crossings in topological semimetal generates the type-I and type-II classification of Dirac/Weyl and nodal-line fermions, accompanied by the exotic electronic and magnetic transport properties.
In this work, we propose a concept of hybrid nodal-chain semimetal, which is identified by the linked type-I and type-II nodal rings in the Brillouin zone. Based on first-principles calculations and swarm-intelligence structure search  technique, a new ternary nitride MgCaN$_2$ crystal is proposed as the first candidate to realize a novel 3D hybrid nodal-chain state. Remarkably, a flat band is emergent as a characteristic signature of such a hybrid nodal-chain along certain direction in the momentum space, thereby serving a platform to explore the interplay between topological semimetal state and flat band. Moreover, the underlying protection mechanism of the hybrid nodal-chain is revealed by calculating the mirror Z$_2$ invariant and developing a $\textbf{\emph{k}}$$\cdot$$\textbf{\emph{p}}$ effective Hamiltonian.
Additionally, considerable drumhead-like surface states with unique connection patterns are illustrated to identify the non-trivial band topology, which may be measured by future experiments.

\end{abstract}

\maketitle

\section{Introduction}

Topological semimetals, emerging as new topological quantum states, have recently attracted vast research interest in both communities of condensed matter physics and materials science\cite{chiu2016classification,armitage2018weyl,Weng2016,wan2011,yu2022encyclopedia}. Thereinto, the nodal line semimetals (NLSMs) with one-dimensional band degeneracies in the momentum space are of remarkable importance due to their intriguing electronic and transport properties\cite{PhysRevLett036806,yu2017topological,yang2018symmetry}. In contrast to 0D nodal point and 2D nodal surface, such 1D nodal line embedded in the 3D Brillouin zone (BZ) is endowed with abundant geometrical and topological structures. Specifically, a single nodal line that subjects to certain crystal symmetry and material parameters may take the forms of an extended straight line\cite{}, a closed ring\cite{yuRRL2015,PhysRevB201114}, or a twisty curve\cite{hang2020double}, demonstrating the morphology of the diversification. More importantly, for the systems with various crystalline symmetries, multiple nodal loops may intersect or interwind with each other, evolving into diverse topological structures such as the nodal-chain\cite{bzduvsek2016nodal,PhysRevLett036401,PhysRevBL081108,PhysRevApplied054080}, nodal-link\cite{PhysRevB.96.041103,PhysRevB.97.155140,PhysRevB.96.041102}, nodal-knot\cite{PhysRevB.96.201305}, and nodal-net\cite{PhysRevLett.120.026402,PhysRevMaterials.2.014202,PhysRevB.102.155116,PhysRevB.98.075146}, providing abundant topological quantum states with exotic properties.

\begin{figure}[t]
\includegraphics[width=8.5 cm]{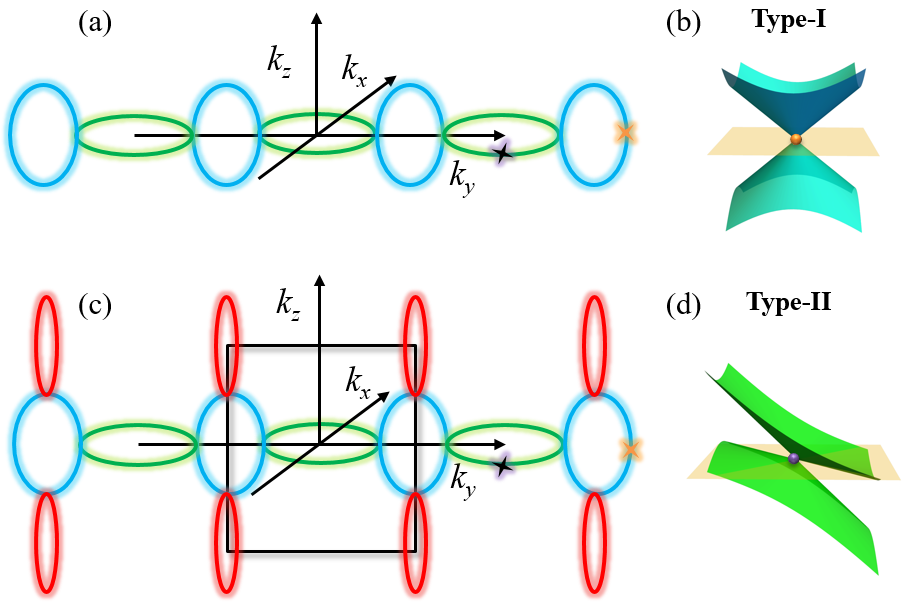}
\caption{Schematic illustrations of (a) 1D nodal-chain and (c) 2D nodal-chain composed of multiple connected nodal rings. The band crossing point on each nodal ring may demonstrate type-I dispersion in (b) or type-II dispersion in (d).}\label{FIG.0}
\end{figure}

Among them, nodal-chain semimetal (NCSM) represents a typical topological structure of multiple NLSMs with distinct topological excitations to achieve non-trivial transport physics\cite{bzduvsek2016nodal,yan2018experimental}.
As shown in Fig. \ref{FIG.0} (a), two nodal rings (represented by blue and green colors) lying on orthogonal planes can interlock together to form a 1D nodal-chain structure extending along the intersecting line, \emph{i.e.} $k_y$ direction. When two sets of 1D nodal-chains couple together, a 2D nodal-chain structure is formed extending along $k_y$ and $k_z$ directions, as seen in Fig. \ref{FIG.0} (c). Such a nodal-chain fermion was initiatively reported by Bzdu\v{s}ek et al\cite{bzduvsek2016nodal} in the non-centrosymmetric and non-symmorphic crystals with multiple glide-plane symmetries in the presence of spin-orbital coupling (SOC). Soon, it was generalized to the spinless systems with SU(2) symmetry without SOC by Takahashi et al.\cite{PhysRevB.96.155206}
In addition, NCSMs were also found to exist in certain centrosymmetric and symmorphic materials\cite{PhysRevB201107,LI2020563,PhysRevB195124,PhysRevB094206,PhysRevB.102.195124,jpcc.7b11075,jpclett.8b02204}, such as WC-type HfC\cite{PhysRevLett036401}, and MgSrSi-type compounds\cite{npj2019-weng}.

Learned from the successfully classification of nodal-point semimetals, a NLSM can also be categorised into different subgroups based on the fold of degeneracy\cite{Fang2016,PhysRevResearch.4.023047} (i.e. two-fold Weyl or four-fold Dirac NLs), the tilting effect\cite{PhysRevB121105,li2017type,he2018type,li2019new,xu2020centrosymmetric,PhysRevB174108} (i.e. type-I, II, or hybrid NLs) and the order of dispersion \cite{PhysRevB.99.121106} (i.e. linear, quadratic, and cubic NLs). These NLSMs can achieve unconventional magnetic responses, such as a zero-field magnetic breakdown and anisotropy in cyclotron resonance\cite{PhysRevB.97.125143,PhysRevLett077202}, thereby inspiring interest on exploring systems with non-trivial band topology.
Motivated by the prosperous of emergent NLSMs, we ask whether a universal classification could be extended into NCSMs, and what kind of related physical phenomena it will bring about.
In fact, from the perspective of degeneracy, a four-fold degenerate ``Dirac nodal chain" was introduced in rhenium dioxide\cite{wang2017nc}, in contrast to previous double-degenerated ``Weyl nodal chain".
Recently,  a ``hybrid nodal-chain" that is composed of $P{\cdot}T$ ($P$: space inversion symmetry, $T$: time reversal symmetry)-protected NLs and mirror-protected NLs was predicted in an orthorhombic graphene network from the perspective of distinctive symmetry\cite{PhysRevBL081108}.
Albeit, the band tilting effect on NCSMs is until now rarely explored and the associated physical mechanism is still uncovered.

In this work, we aim at expanding the category of NCSMs in terms of the band tilting effect. As schematically shown in Fig. \ref{FIG.0}, when two linked nodal-rings belong to the same type-I or type-II case, it can be classified as type-I or type-II NCSM, keeping pace with the category of NLSM.
Unusually, when two nodal-rings inside the NCSM host completely different dispersion types, e.g. one belongs type-I and the other belongs type-II,
it gives birth to a new distinctive nodal-chain state, termed as hybrid NCSM.
In the following, we identify by $\emph{ab initio}$ calculations and structure search method a new ternary nitride MgCaN$_2$ crystal as a novel hybrid NCSM, characterized by the linked type-I and type-II nodal rings around the Fermi level.
The mirror Z$_2$ topological invariants and an effective $\textbf{\emph{k}}$$\cdot$$\textbf{\emph{p}}$ model were adopted to reveal the multi-mirror protection mechanism.
Significantly, a flat-band is emergent as a characteristic signature of such NCSM along certain direction, which serves as a platform for exploring interplay between NCSM states and flat-band physics.
Moreover, the topological surface states with unique connection patterns are demonstrated.
In the final discussion part, we suggest the possible synthesis of MgCaN$_2$ from CaN$_2$ precursor by typical ion implantation strategy, and propose a family of ternary nitride XYN$_2$ (X,Y=Be, Mg, Ca, Sr, Ba) as good candidates for studying distinct NCSM physics.

\section{Computational Methods}
The crystal structure of ternary nitride CaMgN$_2$ was predicted using the crystal structure analysis by particle swarm optimization (CALYPSO) code\cite{PhysRevB094116,WANG20122063}, which has been widely used to search materials with excellent stability and intriguing physical properties.
Structural optimization and electronic structure calculations were carried out by using Vienna $ab$-$initio$ simulation package (VASP)\cite{kresse1996efficiency,kresse1996efficient} with Perdew-Burke-Ernzerhof parameterized generalized gradient approximation (PBE-GGA) \cite{perdew1996generalized}. The energy cutoff of 520 eV and a \emph{k}-point mesh of 12$\times$12$\times$8 are chosen.
The ionic relaxations were performed  until the force on each atom was less than 0.01 eV\verb|/|\AA {} and convergence criterion for the self-consistent electronic minimization loop was set to 10$^{-6}$ eV. Besides, the phonon spectrum was calculated with the PHONOPY package\cite{phonopy}, and the thermal stability was evaluated by performing $ab$-$initio$ molecular dynamics (AIMD) simulations based on an NVT ensemble with temperature controlled by a Nos$\acute{e}$-Hoover bash\cite{Born-Oppenheimer,Nose-Hoover}.
To analyze the electronic topological properties, the tight-binding (TB) Hamiltonian was constructed through projecting the Bloch states into the maximally localized Wannier functions (MLWFs)\cite{mostofi2014updated,marzari2012maximally}, and surface states were calculated using the iterative Green¡¯s function method as implemented in the WannierTools package\cite{wu2018wanniertools}.

\begin{figure*}[t]
\includegraphics[width=17.3 cm]{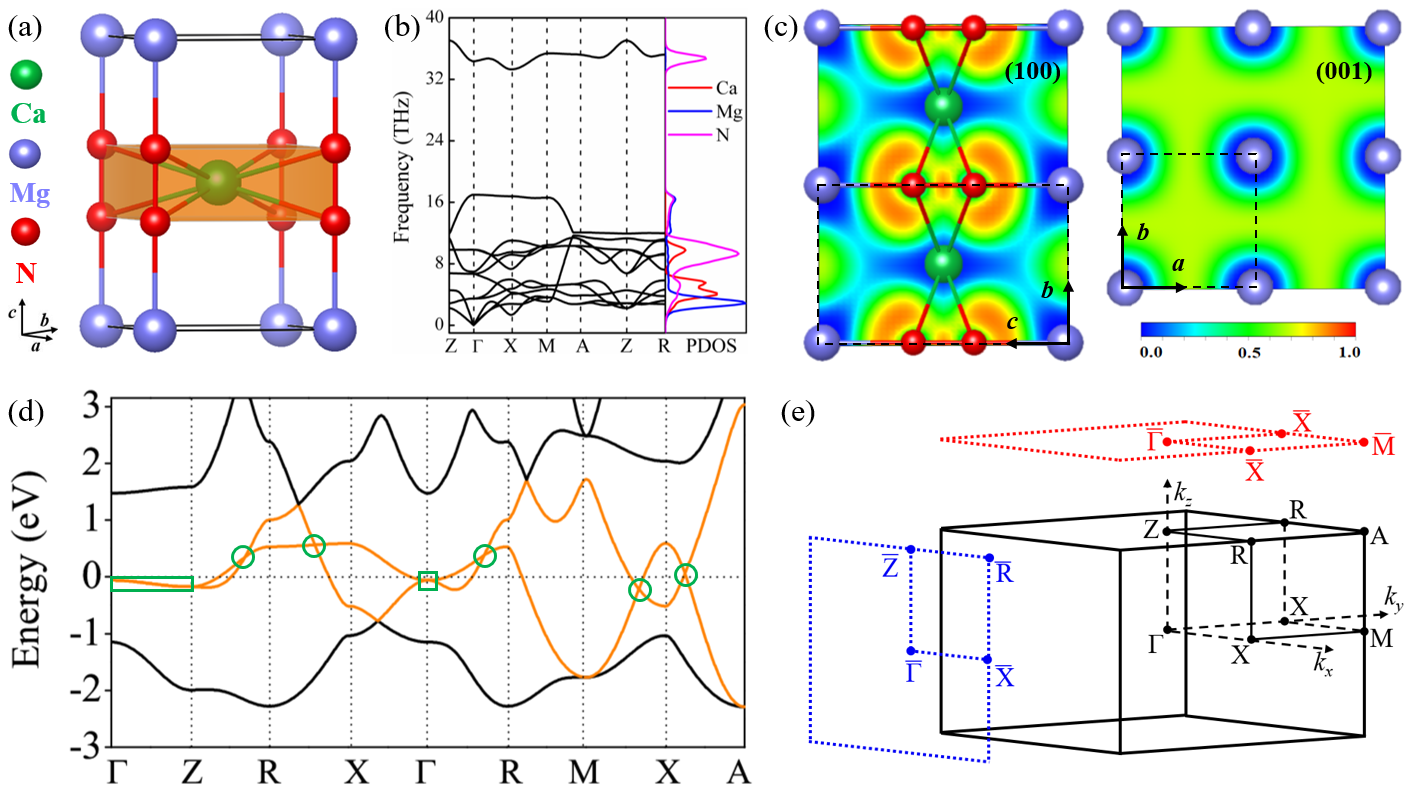}
\caption{(a) The crystal structure of ternary nitride MgCaN$_2$ with optimized lattice constants of $\emph{a}$ = $\emph{b}$ = 3.438 {\AA} and $\emph{c}$ = 5.313 {\AA}. (b) Phonon spectrum and phonon partial density of states (PDOS). (c) The electron localization function (ELF) projected on (100) and (001) surfaces, respectively. (d) The band structure of MgCaN$_2$ at PBE-GGA level. The Fermi level is shifted to zero. (e) The bulk 3D Brillouin zone (BZ) and projected surface BZ of (010) and (001) planes are displayed, and the high-symmetric points are labelled.}\label{FIG.1}
\end{figure*}

\begin{figure*}[t]
\includegraphics[width=18.1 cm]{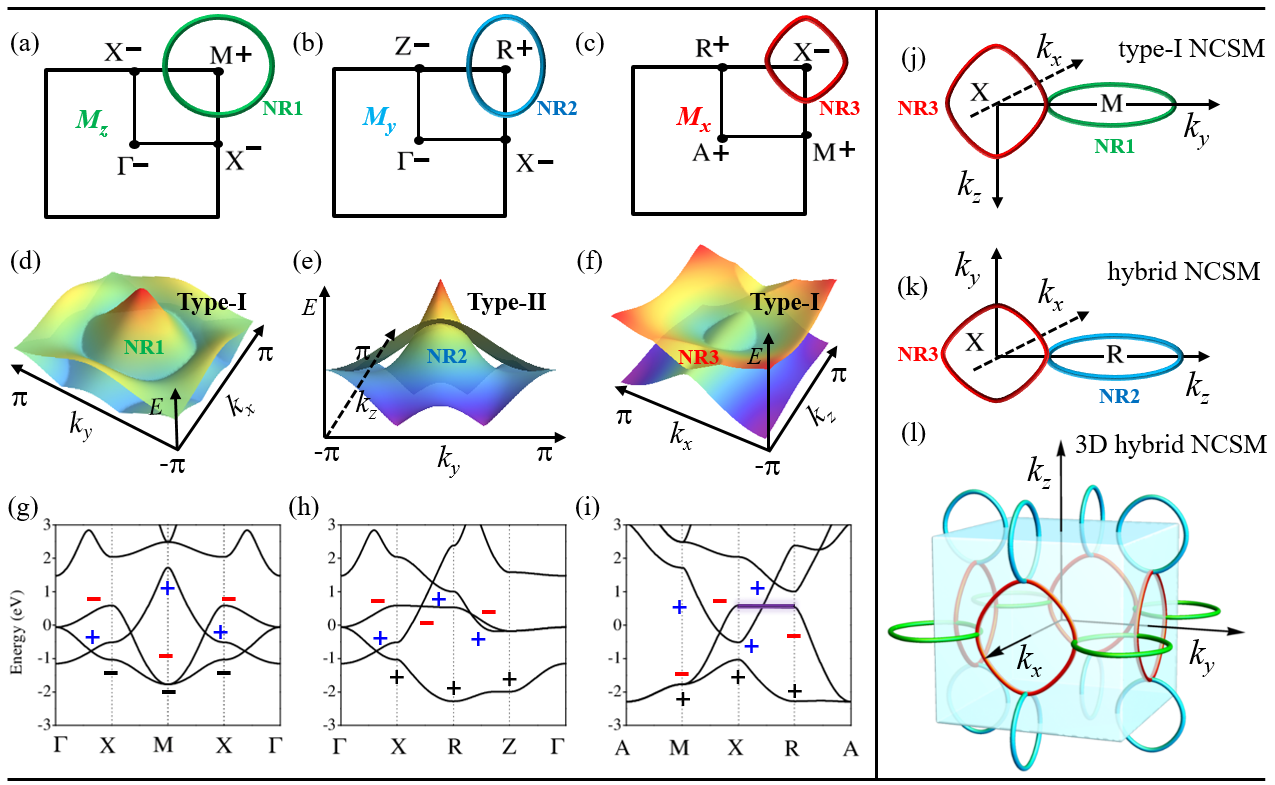}
\caption{(a-c) Diagrams of NR1/NR2/NR3 lying on $M_z$/$M_y$/$M_x$-invariant plane, respectively. The topological index $\delta (\Gamma_m)$ are given on each TRIPs.
(d-f) 3D view of band structures on $k_z$ = 0, $k_y$ = 0 and $k_x$ = $\pi$ planes, respectively.
(g-i) Band structures along high-symmetric paths on corresponding mirror-invariant planes of (a)-(c), where the characteristic flat band of hybrid nodal-chain is highlighted by purple color in (i) along XR path. The associated mirror eigenvalues for all valence bands and the lowest conduction band at TRIPs are marked.
(j-l) Schematically illustrations of 1D type-I nodal chain, 1D hybrid nodal-chain, and 3D hybrid nodal-chain throughout the entire BZ.}\label{FIG.2}
\end{figure*}

\section{Results and anlysis}
\subsection{Crystal structure and stability of MgCaN$_2$}
The crystal structure of ternary nitride MgCaN$_2$ was unveiled by an extensively swarm-intelligence structure search method, which is further optimized based on DFT calculations. As displayed in Fig. \ref{FIG.1} (a), the identified MgCaN$_2$ compound crystallizes in a tetragonal lattice with space group of $P4\verb|/|mmm$ (No. 123). The Ca and N atoms are located at Wyckoff positions 1$d$ (1\verb|/|2, 1\verb|/|2, 1\verb|/|2) and 2$g$ (0, 0, 3\verb|/|8). They form a typical CaN$_2$ layer as observed in CaN$_2$ crystal\cite{schneider2012synthesis}.
The Mg atom intercalates between two layers of CaN$_2$ with Wyckoff position of 1$a$ (0, 0, 0).
The stability of MgCaN$_2$ is firstly investigated by computing the phonon spectrum, as shown in Fig. \ref{FIG.1} (b).
The absence of imaginary-frequency phonon mode throughout the Brillouin zone confirms the dynamical stability of MgCaN$_2$.
Moreover, the low-frequency modes ($\sim$0-6 THz) mainly derive from the Ca and Mg atoms, while the N atoms gain more phonon population in higher frequency modes as expected. In particular, there exist a branch of isolated highest-frequency mode, which is ascribed to the sturdy N-N covalent bond. Besides, the thermal stability of MgCaN$_2$ crystal is also confirmed by AIMD simulations at 300 K and 500 K [see \textcolor[rgb]{0.00,0.07,1.00}{Fig. S1} in Supporting Information (SI)].

The electron localization function (ELF) of MgCaN$_2$ is demonstrated in Fig. \ref{FIG.1}(c). On the (100) plane, there is a dumbbell-shaped iso-value surrounding and inside N-N dimer\cite{liu2012first,kim2018novel}, indicating strong ionic bonding of Ca-N and covalent bonding of N-N, respectively. While on (001) plane crossing the Mg atoms, the electrons tend to diffuse into the interstitial region, manifesting typical metallic bonding nature as that of free electron gas in alkali metal.
In addition, Bader charge analysis gives that each Ca atom loses 1.38 electron ($e$) while each Mg atom only loses 0.96 $e$. All these electrons are transferred to coordinated N atoms. Overall, it reveals that the interactions between Ca-N and Mg-N are both ionic bonding, meanwhile the left electron of Mg$^{+}$ contributes to the metallic bond.

\subsection{Hybrid NCSM state in MgCaN$_2$}
The band structure of MgCaN$_2$ at PBE-GGA level is displayed in Fig. \ref{FIG.1} (d).
We focus on the lowest conduction band and highest valence band (marked by orange color), where several nontrivial band degeneracies emerge around the Fermi level.
First, the band along ${\Gamma}$-Z shows two-fold degeneracy, in which each point exhibits a quadratic dispersion relation in the $k_{x}$-$k_{y}$ plane. This belongs to the quadratic nodal line fermion as proposed by Yu et al\cite{PhysRevB.99.121106}.
Second, various of seemingly isolated band-crossing points exist along Z-R-X, ${\Gamma}$-R, M-X-A paths, which are also checked by using a more advanced hybrid Heyd-Scuseria-Ernzerhof (HSE) functional\cite{HSE1564060} [see Fig. \textcolor[rgb]{0.00,0.00,1.00}{S2} in SI]. Via careful nodal-point searching within the whole Brillouin zone, we reveal that these band crossings actually belong to three kinds of inequivalent nodal rings (NRs), labelled as NR1, NR2, NR3.
As displayed in Fig. \ref{FIG.2} (a-c), the NR1 lies in the horizontal $M_{z}$-invariant plane ($\Gamma$XM) encircling M point, the NR2 lies in the vertical $M_{y}$-invariant plane ($\Gamma$ZRX) centred around R point, and the NR3 lies in the vertical $M_{x}$-invariant plane (ARXM) centred around X point.
Further, 3D band-structures on corresponding mirror-invariant planes are calculated to analyze the band dispersion type of three NRs.
As explicitly demonstrated in Fig. \ref{FIG.2} (d-f), the results show that NR1 and NR3 belong to type-I NRs, whereas NR2 shows type-II feature with over-tilted dispersion relation.

The unique connections for these NRs are unveiled. In Fig. \ref{FIG.2} (j), the type-I NR1 exteriorly contacts with the type-I NR3, to form a type-I nodal-chain periodically elongated along $k_y$ (XM) direction. More intriguingly, as seen in Fig. \ref{FIG.2} (k), the type-II NR2 connects with the type-I NR3, resulting in a hybrid nodal-chain extended along $k_z$ (XR) direction.
Sharing the same NR3, the type-I and hybrid nodal-chains naturally link together to generate a 2D nodal-chain located on $k_x=\pi$ plane.
Further, considering the $C_{4z}$ rotation symmetry of tetragonal lattice, two equivalent 2D nodal-chains on $k_x=\pi$ and $k_y=\pi$ planes have to link together, leading to the realization of a 3D hybrid nodal-chain structure across the entire BZ, as displayed in Fig. \ref{FIG.2} (l).
Therefore, the distinct band tilting of the multiple linked NRs generates a novel hybrid NCSM state in MgCaN2 crystal.

\subsection{Mirror Z$_2$ topological invariant}
To further reveal the underlying formation mechanism of nodal rings and nodal chains in MgCaN$_2$, a Z$_2$ topological invariant derived from the knowledge of the eigenvalues of mirror operator is defined as,
\begin{eqnarray}\label{ti1}
  (-1)^{\nu} = \prod_{m=1}^{4} \delta (\Gamma_m),
\end{eqnarray}
where the sum runs over four time-reversal invariant points (TRIPs) ${\Gamma}_m$ on each mirror-invariant plane.
The quantity $\delta (\Gamma_m)= \prod_{n{\in}occ} {\xi}_{n}^{m} $, in which ${\xi}_{n}^{m}= {\pm 1}$ is the eigenvalue of the $n$th band at ${\Gamma}_m$. The product involves all the occupied bands. In analogy to the well-known parity criterion for topological insulators\cite{FLPhysRevB.76.045302}, here the Z$_2$ invariant $\nu$ = 0 or 1 indicates the existence of odd or even numbers of nodal ring in associated mirror-invariant planes.
Then we apply this criterion to the electronic structure of MgCaN$_2$. The eigenvalues of $M_{x,y,z}$ for Bloch states are calculated and given in Fig. \ref{FIG.2} (g-i).
The fact that two crossing bands around the Fermi level hosting opposite eigenvalues implies the hidden nodal lines on corresponding mirror-invariant planes. Considering all the occupied bands at each TRIP, the $\delta {\Gamma}_m$ is readily obtained and Z$_2$ invariant ${\nu}$ is further calculated, from which one can specifically identify the existence of nodal rings with distinctive shapes.

For instance, as for the $M_x$-invariant plane (ARXM) in Fig. \ref{FIG.2} (c), the ${\delta}$(R) = ${\delta}$(A) = ${\delta}$(M) = $-{\delta}$(X) = $+$1 leads to $\nu$ = 1.
Hence, there must be at least a nodal line inside this plane. More interesting, the opposite ${\delta}$ between X and R/A/M suggests that
the band inversion should happen for odd times along arbitrary $k$-path connecting X and the other points, which directly leads to the emergence of a closed nodal ring surrounding X point, namely NR3. Analogously, in Fig. \ref{FIG.2} (a-b) the associated topological indexes on $\Gamma$XM and $\Gamma$ZRX ensure the appearance of NR1 centered at M point and NR2 centered at R point, respectively.
Because the XM belongs to the intersection of $\Gamma$XM and ARXM planes, such that the band crossing point at XM belongs to NR1 and NR3, simultaneously. As a result, the X-centered NR1 and M-centered NR3 have to interlock with each other exactly at the nexus point\cite{xiong2020hidden}, forming a 1D nodal-chain along XM direction. Similar analysis works for XR path as the intersection of $\Gamma$ZRX and ARXM planes, giving anther 1D nodal-chain along XR direction. These results derived from topological indexes are qualitatively in agreement with results from first-principle calculations.


\begin{table*}
\centering
\caption {Fitting parameters $w_i$ and $m_i$ of $k$$\cdot$$p$ Hamiltonian around ${\Gamma}_m$ for each nodal ring on the relevant mirror-invariant plane in the unit of eV.}
\setlength{\tabcolsep}{4 mm}
\renewcommand\arraystretch{1}
\tabcolsep=0.5 cm
{
\begin{tabular}{ccccccccccc}
  \toprule
   plane & ${\Gamma}_m$ & $w_0$ & $w_x$ & $w_y$ & $w_z$ & $m_0$ & $m_x$ & $m_y$ & $m_z$ \\
    \hline
  $\Gamma$XM & M & -0.024 & 0.420 & 0.420 & $-$ & 1.744 & 4.920 & 4.920 & $-$ \\
    \hline
  $\Gamma$ZRX & R & 0.024 & 2.520 &$-$ & -3.660 & 0.550 & 5.690 & $-$ & 3.720\\
    \hline
  ARXM & X & 0.775 &$-$ & 5.170 & 0.776 & 0.225 & $-$ & 2.960  & 0.800  \\
    \hline
\end{tabular}
}
\label{table1}
\end{table*}

\subsection{Effective $\textbf{\emph{k}}$$\cdot$$\textbf{\emph{p}}$ Hamiltonian and inevitable flat band}
To obtain quantitatively understanding of these linked nodal rings, particularly the distinctive dispersion relations, we now construct the low-energy effective $k$$\cdot$$p$ model for each nodal ring on corresponding mirror-invariant plane.
In general, a two-band $\textbf{\emph{k}}$$\cdot$$\textbf{\emph{p}}$ Hamiltonian around the $\Gamma_m$ in BZ can be written as:
\begin{equation} \label{kp1}
H_{\Gamma_m}(\textbf{q})=\sum _{i=0,x,y,z} f_i(\textbf{q})\sigma_i
\end{equation}
where $f_{0,x,y,z}(\textbf{q})$ are real functions, and $\textbf{q}=(q_x, q_y, q_z)$ is the momentum vector respect to the ${\Gamma}_m$. $\sigma_0$ is the identity matrix, and $\sigma$$_{x,y,z}$ denotes the Pauli matrices.

Now, we consider the symmetry constrains of the Hamiltonian in Eq. \ref{kp1}. In the absence of SOC, the time-reversal symmetry can be taken as $T=K$, where $K$ is the complex conjugation operator.
As the $TH(\textbf{q})T^{-1}=H(-\textbf{q})$ requires the $f_y(\textbf{q})$ and $f_{0,x,z}(\textbf{q})$ to be odd and even functions, respectively.
As for the $M_{i=x,y,z }$-invariant plane, based on the fact that two bands have opposite mirror eigenvalues, then $M_{i}$ can be written as $M_{i}={\sigma}_z$, which further adds constraint $f_{x,y}(\textbf{q})$ and $f_{0,z}(\textbf{q})$ to be odd and even functions, respectively. Hence, in the $M_{i }$-invariant plane, up to the second order of $\textbf{q}$, the model Hamiltonian is
\begin{equation} \label{kp-mads}
H_{\Gamma_m}(\textbf{q})|_{q_i=0}= f_0(\textbf{q})\sigma_0+ f_z(\textbf{q})\sigma_z,
\end{equation}
with
\begin{equation} \label{kpf0}
f_0(\textbf{q})=w_{0} - w_{j}q_{j}^{2}-w_{k}q_{k}^{2},
\end{equation}
and
\begin{equation} \label{kpfz}
f_z(\textbf{q})=m_{0} - m_{j}q_{j}^{2}-m_{k}q_{k}^{2}.
\end{equation}
Here, $i,j,k=x,y,z$, $i\neq j\neq k$, the parameters $w_i$ and $m_i$ $(i=0, x, y, z)$ are derived by fitting the band dispersions of model Hamiltonian with that of DFT calculations, as summarized in Table \ref{table1}.

The shape of the nodal ring in $M_{i }$-invariant plane is determined by the solution of $f_{z}(\textbf{q})|_{q_i=0}=0$.
Taking the $M_z$ symmetry as an example, on its invariant plane $\Gamma$XM, an effective Hamiltonian $H_M$ is built centered on M point.
We readily get $m_{0}-m_{x}q_{x}^{2}-m_{y}q_{y}^{2} = 0$.
Apparently, the signs of $m_{0, x, y}$ determine the existence and the shape of the nodal ring, e.g. the $m_{0,x,y}>0$ refers to a closed nodal-loop encircling M point, while the $m_{0,x}>0$ and $m_{y}<0$ leads to two opened nodal-loops traversing the BZ.
Here,  as tabulated in Table  \ref{table1}, the parameters in  $H_\mathrm{M}$ are fitted to be $m_0=1.744 > 0$ and $m_{x,y}=4.920>0$, which naturally implies a closed loop encircling M, namely NR1.
Likewise, Hamiltonian $H_\mathrm{R}$ with $m_0>0$, $m_{x,z}>0$ guarantees the NR2 on ${\Gamma}$ZRX plane, and Hamiltonian $H_\mathrm{X}$ with $m_0>0$, $m_{y,z}>0$ determines the NR3 on ARXM plane.

\begin{figure}[t]
\includegraphics[width=8.8 cm]{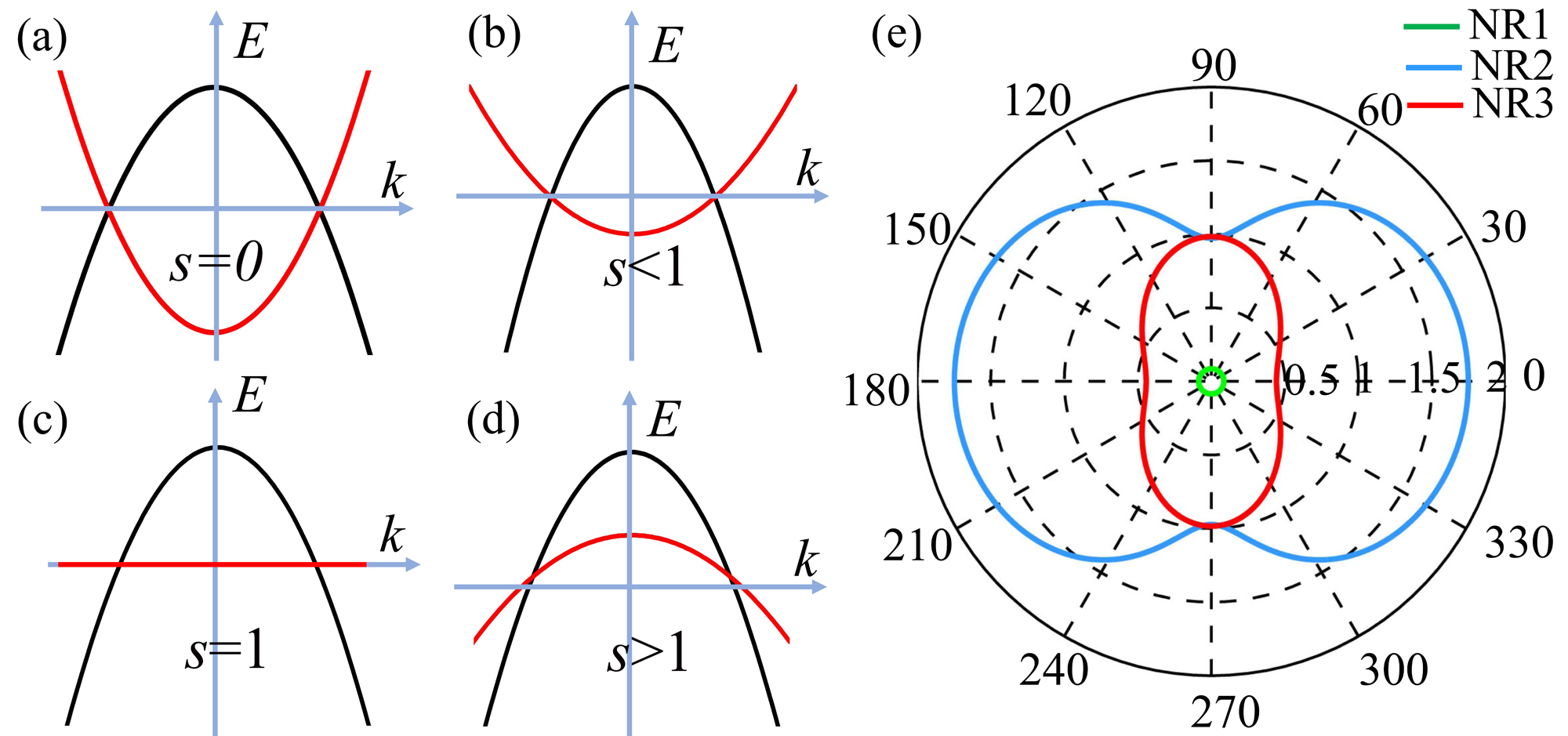}
\caption{(a)-(d) Demonstrations of distinctive band titling dominated by $s(\theta)$. (e) The computed $s(\theta)$ values for the NR1 (green line), NR2 (blue line), and NR3 (red line), respectively.}\label{Fig-kp}
\end{figure}

\begin{figure*}[t]
\includegraphics[width=17.3 cm]{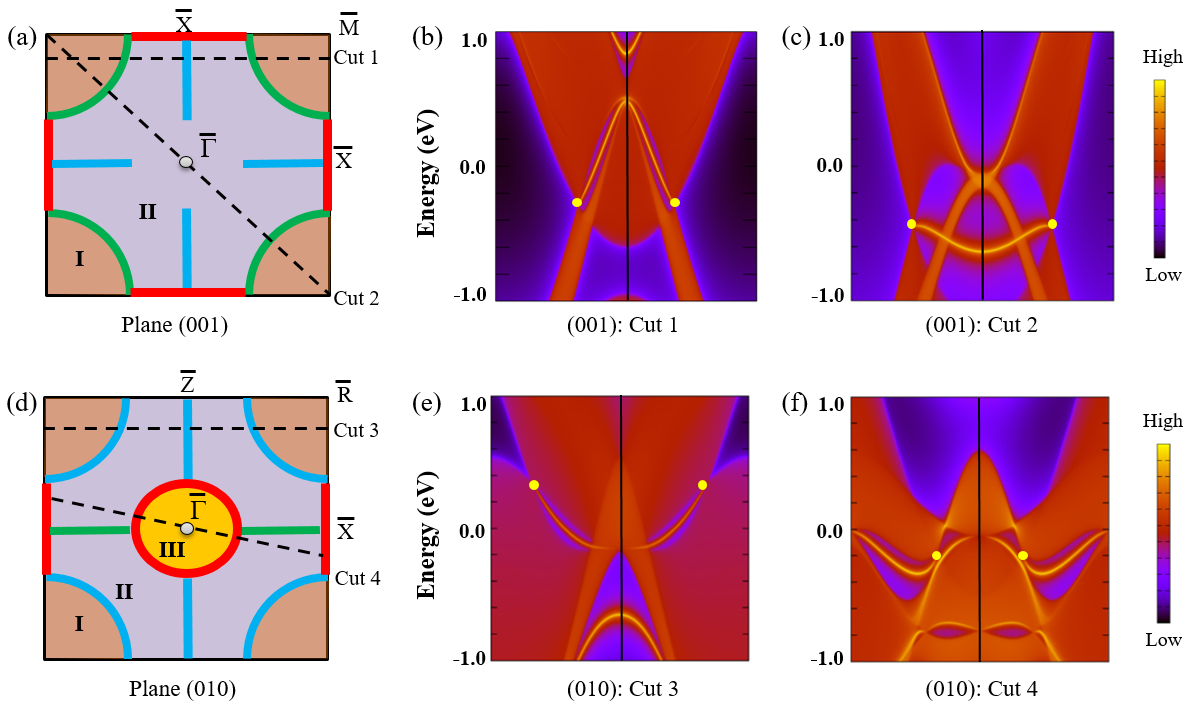}
\caption{Diagrams of the 3D hybrid nodal chain projected onto surface BZ of (a) (001) and (d) (010) planes. (b-c) and (e-f) show the calculated topological surface states of MgCaN$_2$ along different cuts as labeled by the dashed lines in (a) and (d).}\label{FIG.3}
\end{figure*}

Applying the replacements of $q_{j}$ by $\left|q\right|\mathrm{cos}\theta$ and $q_{k}$ by $\left|q\right|\mathrm{sin}\theta$. The eigenvalues of Hamiltonian Eq. \ref{kp-mads} are,
\begin{eqnarray}
E_{\pm} & = & f_{0}(\textbf{q})|_{q_{i}=0}\pm f_{z}(\textbf{q})|_{q_{i}=0}\nonumber \\
 & = & \boldsymbol{\epsilon}_{0}-\textbf{q}^{2}\left(\boldsymbol{w}\pm\boldsymbol{m}\right),
\end{eqnarray}
with $\epsilon_{0}=w_{0}+m_{0}$, $\boldsymbol{w}=w_j  \textup{cos}^2\theta +   w_k \textup{sin}^2\theta$, and $\boldsymbol{m}=m_j  \textup{cos}^2\theta +   m_k \textup{sin}^2\theta$.
Hence, the dispersion relation (type-I or type-II) of NRs in  $M_{i }$-invariant planes can be estimated by a direction-dependent parameter $s(\theta)$, which is defined as:
\begin{equation}
s(\theta)=\frac{ \left| w_j \right| \textup{cos}^2\theta +  \left| w_k \right|\textup{sin}^2\theta }{\left| m_j \right| \textup{cos}^2\theta +  \left| m_k \right| \textup{sin}^2\theta }.
\end{equation}
The nonzero $s(\theta)$ can break the electron-hole symmetry, leading to the asymmetric nature of conduction and valence bands around the $\Gamma_m$ point.
It allows us to roughly estimate the types of dispersion for the crossing points.
Specifically, for $s(\theta)<1$, this case indicates along $\textbf{q} (\theta)$ direction the parabolic conduction and valence bands have opposite opening direction, thus leading to type-I band crossing point as shown in Figs. \ref{Fig-kp}(a)-(b).
In contrast, $s(\theta)>1$ indicates the conduction and valence bands have same opening direction, thus generating type-II band crossing points as shown in Fig. \ref{Fig-kp}(d).
It's worth mentioning that for $s(\theta) = 1$, this special case implies a flat band along certain direction, generating unique type-III band crossing\cite{PhysRevB.101.100303}, as displayed in Fig. \ref{Fig-kp}(c).

The values of $s(\theta)$ for target NR1/NR2/NR3 are given in Fig. \ref{Fig-kp} (e).
We find that $s(\theta)<<1$ for NR1 (the circle in green color) along any directions, demonstrating that NR1 is a typical type-I nodal ring.
For NR2 (blue color), the condition $s(\theta)>1$ is satisfied for most of $\theta$ aside from $\theta= 90^0$ or $180^0$ (along RX), where the $s=0.97 \approx 1$.
Thus, NR2 can be approximatively taken as a type-II nodal ring.
As for NR3 (red color), one can find $s(\theta)<1$ always hosts for arbitrary direction, which indicates that NR3 belongs to type-I.
Specially, the $s(90^0) = 0.98 \approx 1$, implies the emergence of flat band along XR direction, as confirmed by the DFT result in Fig. \ref{FIG.2}(i). Significantly, the rise of flat band is dominant characteristic of the hybrid NCSM.
Since a hybrid nodal-chain state is formed by connected type-I nodal ring centering on $\Gamma_1$ and type-II nodal ring centering on $\Gamma_2$, which require $s\leq1$ and $s\geq1$  along $\Gamma_1 \Gamma_2$, simultaneously. Thus, the $s=1$ is restricted along $\Gamma_1 \Gamma_2$, which inevitable results in a flat band. This is exactly the origin of the flat band along the intersectional XR in MgCaN$_2$.
The interplay between the nontrivial topology and flat band in hybrid NCSM may provide a promising route towards unconventional superconductivity and correlated electronic states\cite{regnault2022catalogue,PhysRevX.11.031017,volovik2013flat}.

\subsection{Topological surface states}
NCSMs usually possess distinctive surface states on different direction.
By using the post-processing code of WannierTools\cite{wu2018wanniertools}, we have computed the (001) and (010) surface states for MgCaN$_2$ as shown in Fig. \ref{FIG.3}.
For the (001) surface in Fig. \ref{FIG.3}(a), the NR1 projects into a closed ring (green circle) encircling $\overline{\rm{M}}$, while the NR3 projects into a line segment (red line), which connects NR1 with its periodic repeated unit. Besides, the NR2 is also projected into a line segment (blue line) connecting NR3 at $\overline{\rm{X}}$. Those projected nodal rings divide (001) surface into two separated regions: region-I, projection of NR1 which is a closed isolated region; region-II, outside projection area of NR1, which is an extended region that stretches across the entire BZ. From surface spectra along two typical paths, \emph{i.e.} cut 1 in Fig. \ref{FIG.3}(b) and cut 2 in Fig. \ref{FIG.3}(c),
we discover that the typical drumhead surface state\cite{PhysRevB.92.045108} only emerges in the region-II, and would be pinned by the projection of NR2.

For the (010) surface in Fig. \ref{FIG.3} (d), the NR2 on $k_y=0$ plane projects into a closed ring (blue circle) encircling $\overline{\rm{R}}$, while the NR3 on $k_y=\pi$ plane projects into a closed ring (red circle) encircling $\overline{\Gamma}$, which divides (010) surface into three separated regions of I, II and III.
Meaningwhile, the equivalent NR2 from $k_x=0$ plane projects in to a line segment (blue line) along $\overline{\Gamma}\overline{\rm{Z}}$ which connects projected NR3 (red ring), while the equivalent NR3 on $k_x=\pi$ plane projects into a line segment (red line) along $\overline{\rm{X}}\overline{\rm{R}}$ which connects the projected NR2 (blue ring).
Besides, the NR1 is also projected into a line segment (green line) along $\overline{\Gamma}\overline{\rm{X}}$, which connects the projected NR3.
The surface spectra along two typical paths are displayed, cut 3 in Fig. \ref{FIG.3}(e) and cut 4 in Fig. \ref{FIG.3}(f).
For cut 3, which passes through region-I and -II, we find the drumhead surface state from the type-II Dirac point of NR2 passes through the region-II, and is pinned by the projected NR2 on $\overline{\Gamma}\overline{\rm{Z}}$.
For cut 4, which passes through region-II and -III, we discover that the drumhead surface state from the projected NR3 (red ring and line) only exists inside the region-II. Such rich and extended surface states of MgCaN$_2$ in the momentum space are feasibly measured by surface experimental techniques, such as scanning tunneling microscope or angle-resolved photoemission spectroscopy\cite{belopolski2019discovery,hosen2020experimental,PhysRevX.10.011026}.

\begin{figure}[t]
\includegraphics[width=8.7 cm]{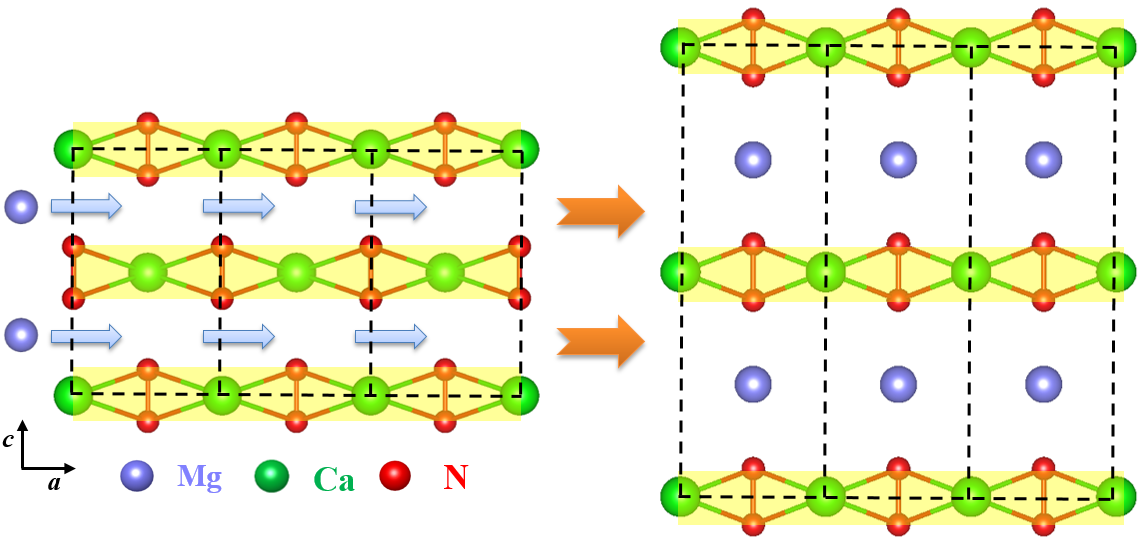}
\caption{Schematic illustrations of the possible synthesis of MgCaN$_2$ via the Mg ion intercalating into CaN$_2$ crystal.}\label{FIG.4}
\end{figure}


\section{Discussions and conclusions}
Though ternary nitride MgCaN$_2$ is discovered from the structure search method, we propose a possible synthetic route of MgCaN$_2$ compound by the powerful technique of ion intercalation\cite{nikitina2017transport,zhao2020engineering,sood2021electrochemical}.
As an typical alkaline earth diazenides, CaN$_2$ crystallizes\cite{schneider2012synthesis} in a tetragonal lattice with space group of $I4/mmm$ (139), as shown in Fig. \ref{FIG.4}. It can be regarded as AB-stacking of CaN$_2$ layers along $c$ direction.
By interpolating Mg ion into the interstitial region between two adjacent CaN$_2$ layers, the interlayer distance is significantly enlarged in accompany with interlayer sliding, which leads to the formation of a new phase MgCaN$_2$.
With unique ionic-bond between Mg$^{+}$ and [CaN$_2$]$^{-}$ and covalent-bond within [CaN$_2$]$^{-}$, ternary nitride MgCaN$_2$ belongs famous Zintl compounds, such as CaAl$_2$Si$_2$\cite{kauzlarich2016zintl,zheng1986site}.
Moreover, a group of isoelectronic compounds with chemical formula XYN$_2$ (X, Y = Be, Mg, Ca, Sr, Ba) are designed as illustrated in Fig. \textcolor[rgb]{0.00,0.00,1.00}{ S3}-\textcolor[rgb]{0.00,0.00,1.00}{S4} on SI. These materials are revealed to host abundant topological semimetals states including NLSM and NCSM with distinctive dispersion types as summarized in Table \textcolor[rgb]{0.00,0.00,1.00}{S1} in SI.


In summary, based on first-principles calculations, swarm-intelligence structure search method, and symmetry analysis, we predicted a new Zintl compound MgCaN$_2$ as a hybrid NCSM, with interlocked type-I and type-II nodal rings around the Fermi level. The confirmed good dynamical and thermal stability demonstrates its promise for experimental synthesis.
By analyzing the mirror Z$_2$ topological invariant and developing a $\textbf{\emph{k}}$$\cdot$$\textbf{\emph{p}}$ model, we unveiled the physical origin of the hybrid nodal-chain structure with multiple-mirror protected mechanism.
More importantly, we pointed out that a flat band along a certain direction is emergent as a characteristic signature of the hybrid nodal-chain, thus serving as a platform to explore the interplay between the topological semimetal state and flat band.
Additionally, the considerable drumhead-like surface states were observed on the (010) and (001) surfaces, benefiting the experimental detection of the nontrivial band topology.
Our work expands the understanding of nodal-chain semimetals from the perspective of distinct band tilting effects, and provides an ideal platform to realize novel hybrid nodal-chain state for further experimental investigations.

\begin{acknowledgements}

This work is supported by the Science and Technology Project of Hebei Education Department (Grant No. BJ2021028), the Natural Science Foundation of Hebei Province (Grant No. A2022205014), and the PhD research startup foundation of Hebei Normal University (Grant No. L2021B11). Dr. D.-S. Ma acknowledges the funding from the China National Postdoctoral Program for Innovative Talent (Grant No. BX20220367) and National Natural Science Foundation of China  (Grant No. 12204074). Dr. B. Fu thanks the National Natural Science Foundation of China (Grants No. 12204330) and the Sichuan Normal University for financial support (Grant No. 341829001).

\end{acknowledgements}

\end{document}